\documentclass[aps,twocolumn,showpacs,nofootinbib,showkeys,superscriptaddress,preprintnumbers,longbibliography,10pt]{revtex4-2}
\usepackage[utf8]{inputenc}

\usepackage{latexsym}
\usepackage{epsfig}
\usepackage{graphicx,subfigure}
\usepackage[normalem]{ulem}
\usepackage{color}
\usepackage{xcolor}
\usepackage{multirow}
\usepackage{hyperref}
\usepackage{numprint}
\usepackage{eucal}
\usepackage{amsmath}
\usepackage{amssymb}
\usepackage{amsmath,amssymb} 
\usepackage{wasysym}

\usepackage{bm}

\DeclareFontFamily{U}{mathb}{\hyphenchar\font45}
\DeclareFontShape{U}{mathb}{m}{n}{
      <5> <6> <7> <8> <9> <10> gen * mathb
      <10.95> mathb10 <12> <14.4> <17.28> <20.74> <24.88> mathb12
      }{}

\definecolor{blue}{rgb}{0,120,250}

\usepackage{xcolor}

\newcommand{\ssout}[1]{}
\begin{document}

\def\ga{\mathrel{\raise.3ex\hbox{$>$\kern-.75em\lower1ex\hbox{$\sim$}}}}
\def\la{\mathrel{\raise.3ex\hbox{$<$\kern-.75em\lower1ex\hbox{$\sim$}}}}

\def\be{\begin{equation}}
\def\ee{\end{equation}}
\def\bea{\begin{eqnarray}}
\def\eea{\end{eqnarray}}

\def\betap{\tilde\beta}
\def\del{\delta_{\rm PBH}^{\rm local}}
\def\Msun{M_\odot}

\newcommand{\dd}{\mathrm{d}} 
\newcommand{\Mpl}{M_P} 
\newcommand{\mpl}{m_\mathrm{pl}} 

\newcommand{\CHECK}[1]{{\color{red}~\textsf{#1}}}

\title{Searching for large dark matter clumps using the Galileo Satnav clock variations}

\author{Bruno Bertrand}
\affiliation{Royal Observatory of Belgium, Avenue Circulaire, 3, 1180 Uccle, Belgium}

\author{Pascale Defraigne}
\affiliation{Royal Observatory of Belgium, Avenue Circulaire, 3, 1180 Uccle, Belgium}

\author{Aurélien Hees}
\affiliation{SYRTE, Observatoire de Paris-PSL, CNRS, Sorbonne Université, LNE, 61 avenue de l’Observatoire, 75014 Paris, France}

\author{Alexandra Sheremet}
\affiliation{SYRTE, Observatoire de Paris-PSL, CNRS, Sorbonne Université, LNE, 61 avenue de l’Observatoire, 75014 Paris, France}

\author{Clément Courde}
\affiliation{Université Côte d'Azur, CNRS, Observatoire de la Côte d'Azur, IRD, Géoazur, 2130 Route de l'Observatoire, 06460 Caussols, France}

\author{Julien Chabé}
\affiliation{Université Côte d'Azur, CNRS, Observatoire de la Côte d'Azur, IRD, Géoazur, 2130 Route de l'Observatoire, 06460 Caussols, France}

\author{Javier Ventura-Traveset}
\affiliation{European Space Agency, 18 Av. Edouard Belin, 31400 Toulouse, France}

\author{Florian Dilssner}
\affiliation{European Space Agency, Robert-Bosch-Straße 5, 64293 Darmstadt, Germany}

\author{Erik Schoenemann}
\affiliation{European Space Agency, Robert-Bosch-Straße 5, 64293 Darmstadt, Germany}

\author{Luis Mendes}
\affiliation{European Space Agency, Camino Bajo del Castillo s/n, 28692 Villafranca del Castillo, Spain}

\author{Pacôme Delva}
\affiliation{SYRTE, Observatoire de Paris-PSL, CNRS, Sorbonne Université, LNE, 61 avenue de l’Observatoire, 75014 Paris, France}

\begin{abstract}
This study presents bounds on transient variations of fundamental constants, with typical timescales ranging from minutes to months, using clocks in space.
The underlying phenomenology describing such transient variations relies on models for Dark Matter (DM) which suggest possible encounters of macroscopic compact objects with the Earth, due to the motion of the solar system in the galactic halo. If such compact objects possess an effective feeble interaction with the ordinary matter beyond the gravitational one, it may result in effective transient variations of fundamental constants. Such variations leave signatures on clocks onboard GNSS satellites. In this paper, we introduce a phenomenological study dedicated to the search for such DM transient objects using the network of passive hydrogen masers (H-Masers) onboard Galileo satellites.  
We first model the signature of transient variations of fundamental constants as a frequency modulation in the difference between two satellite clocks, considering the satellite trajectories relative to the transient event. Then, we present first results based on a fast analysis method, the maximum reach analysis. The main result is a significant extension of the discovery range for DM transients, with a sensitivity never achieved before. We investigate indeed the range of transient sizes from $10^5$ to $10^9$ kilometres, which, apart from indirect and model-dependent non-transient effects, has never been explored previously.


%
\end{abstract}

\maketitle




\setcounter{footnote}{0}



\section{Introduction}

\subsection{Varying fundamental constants}

How may a constant vary? This question may sound semantically absurd. However, it turns out to be of prime importance in the quest of unknown physics beyond the current theoretical models. Indeed, fundamental constants are free parameters intrinsic to the theory that introduces them. Consequently, a physical quantity may be qualified as constant in space and time  within one given theoretical framework, e.g. the Standard Model (SM) of particle physics and General Relativity (GR). Therefore, a hypothetical variation of physical quantities which are supposed to be constant in the SM or the GR could reveal some failure of these models, or equivalently, could be the discovery of new physics beyond these models, as introduced in \cite{Damour:2010rp,Uzan:2010pm}.
Currently, the SM is described by 22 fundamental constants determined by experiments. For example, the reduced Planck constant $\hbar$, the elementary charge $e$, the speed of light $c$, nine Yukawa couplings for leptons and quarks, etc. From these former are derived the rest mass of particles like $m_\textrm{p}$ (proton), $m_\textrm{e}$ (electron)... Their absolute value depends on a given system of units. However, in practice, a physical measurement of these constants is a comparison of two quantities. The measured  Subsequently, dimensionless combinations of constants are measured, called ``fundamental parameters''.

The purpose of our project is to exploit the capabilities of atomic clocks onboard the Galileo satellites, to search for transient variations of fundamental parameters. In this paper, we focus only on the fine structure constant and the proton-to-electron mass ratio. Notice that other fundamental parameters can also be tested, based for example on the light quarks mass, see \cite{Damour:2010rp}. The fine structure constant, denoted by $\alpha$, governs the strength of the electromagnetic interaction and is  almost equal to 1/137. 
The proton-to-electron mass ratio is denoted by $\mu$ such that:
\begin{equation}\label{def:mu_ep}
 \mu = \frac{m_\textrm{p}}{m_\textrm{e}}\approx 2000 \, .
\end{equation}
Several astrophysical observations suggest that ordinary matter contributes only to around 5\% to the total mass-energy content of our Universe. The unknown remaining part is commonly separated into the dark matter (DM) and the dark energy. Some theoretical models suggest that DM consists of a massive scalar field of which the feeble interaction with ordinary matter produces space-time variations of fundamental constants (see \cite{Stadnik:2015kia,Arvanitaki:2014faa,Hees:2018fpg} and references therein). 

Recently, a family of such DM models which describes clusters or classical structures of macroscopic size, called DM clumps, have gained considerable interest. They are viable DM candidates that could regularly cross the Earth, hence the name of ``DM transients'', and consist in the condensation of the scalar field into spatially extended objects like, e.g., the topological defects (\cite{Derevianko:2013oaa,Pospelov:2012mt}), nontopological solitons (\cite{Kusenko:2001vu}), dark clusters (\cite{Dror:2017gjq,Visinelli:2018wza}) or gravitationally bound particles into dark stars (\cite{Eby:2015hsq,Braaten:2018nag,Banerjee:2019epw}).
During such hypothetical encounters, it has been shown in \cite{Derevianko:2013oaa,Pospelov:2012mt} that a variation of fundamental constants is possibly observable by means of precision measurement devices, like atomic clocks which would experience some frequency change. In particular, if the trajectory of the Earth in the Galactic halo intercepts such a (large enough) DM transient, a geographically distributed network of atomic clocks, like the one offered by the GNSS constellations, is expected to exhibit a specific pattern when the clock relative frequencies are compared. This approach can be compared to the gravitational wave detection, where correlations between detectors are a strong indicator of a real event.

However, some parameters inherent to these models are poorly constrained for the moment: 1) the mean duration $\mathcal{T}$ between two consecutive encounters of transients with the Earth, 2) the size $d$ of the transients, 3) the energy scale $\Lambda_\textrm{x}$ governing the interaction strength between the DM and the ordinary matter. The 3-dimensional volume in which the DM transients can be detectable in adequacy with a combination of these three parameters is called the parameter space. The ultimate goal is the direct detection of DM, but the lack of experimental observations, in regions of parameter space where transient objects may exist, means that certain models can be excluded.

\subsection{The GASTON Project}

One of the aims of the GASTON (GAlileo Survey of Transient Objects Network) project is to exploit the extremely stable H-maser clocks onboard satellites of the Galileo constellation to search for signatures of transient variations of fundamental parameters in the close vicinity of the Earth. 
%
Another point is that our modelling is based on a specific scalar field Lagrangian, which leads to a smooth profile for transient DM clumps, and a subsequent smooth profile for the variation of fundamental constants.
In addition, we also consider for the first time the orbital motion of the satellites. This enables to extend the potential detection to any large DM objects, well beyond the size of the Earth, namely a significant extension of the current literature dealing with the GPS constellation \cite{Derevianko:2013oaa,Roberts:2017hla} or optical atomic clocks \cite{Wcislo:2018ojh}. 

The GASTON project can be divided into two phases. The first one is the ``exclusion approach'' which consists in excluding large regions of the parameter space through a first rough analysis. That's the main purpose of the present work as described in Section \ref{sec:Max-reach_analysis} which will deal with a simple and fast ``maximum reach analysis'' (MRA). 
The advantage of such an exclusion is to narrow down the specific region of parameter space in which to search for potential signatures of DM transients. 
The second phase is called “detection approach” and will be the subject of a future paper. It deals with an exhaustive correlation analysis over the whole set of available Galileo satellite clocks to look for possible signatures of variation of fundamental constants, or, if not, to refine the exclusion zone. 

\subsection{This work: Strong point, approximations and limitations}
\label{sec:This-work}

Domain walls (DW), a special case of topological defects with a planar configuration, have been introduced by \cite{Pospelov:2012mt,Derevianko:2013oaa} as a DM candidate. This papers suggested that a network of domain walls with a macroscopic thickness between $10^{-1}$ and $10^4$ km could at least partially fulfill the local cold DM energy density. In addition, \cite{Roberts:2017hla} proposes a DM model of bubbles that satisfies the cosmological equation of state for DM. These bubbles are formed by domain walls closing on themselves and induce a signature similar to the planar DW when their transverse size significantly exceeds the terrestrial scale. However, it was discussed in \cite{Stadnik:2020bfk} that minimal models of DM topological defects crossing frequently the Earth trajectory are not compatible with standard cosmological models. Moreover, depending on the sign of the coupling constant, such a topological defect might be affected by the repulsive or attractive potential generated by the Earth \cite{Stadnik:2020bfk,stadnik:arXiv2021, derevianko:arXiv2021}. This can lead to a screening or an enhancement of the signal searched for.

However, numerous models allow to consider possible encounters between Earth and DM clumps, characterized by a feeble scalar interaction with ordinary matter in addition to the gravitational interaction. Hence, in this paper, we use domain walls as a toy model for DM clumps based on the condensation of a scalar field. We describe two cases of ``thin'' and ``large'' domain walls. Our relative definition of ``thin and large'' refers to the comparison of the typical DW size with the typical distance between satellites  (see Section \ref{sec:Distance Threshold} for more details). 
(1) The ``thin'' domain wall case represents a thin planar clump or a ``large'' clump (regardless of its shape) with a skin effect, where the variation of the scalar field is confined to a thin layer below the surface. In such a case, the typical signature at the level of the clock satellite is either a frequency spike or an over-damped frequency periodic oscillation.
(2) The ``large'' domain wall case represents a large (e.g. spherical) DM clump with a smooth radial profile from the center to the edge. Throughout the rest of this paper, we will refer to these as ``large transients''. The typical signature induced by such transients on the clock satellite is a frequency modulation at the orbital period, with a modulation depending on the DM interaction and the scalar field profile. This limit is addressed and modeled for the first time in this manuscript.

In addition to using precise onboard passive H-Masers with the Galileo constellation, our approach presents several advantages:
\begin{itemize}
    \item 
    It can be easily generalised to various models of DM transient.
    \item As previously mentioned in case of self-interacting double-well potential, the Earth can produce a (anti-)screening effect which can generate an effective potential with a single minimum, according to the sign of the coupling constant. This is due to the back reaction caused by the large density of baryonic and leptonic matter from the Earth, which reduces or enhances the sensitivity of experiments conducted on the Earth. Using clocks in space mitigates the first problem, since the back reaction is mainly restricted to the satellite mass density.
    \item If we consider thick domain walls not as a toy model but as a realistic model as in \cite{Roberts:2017hla,Stadnik:2020bfk}, a full statistical analysis provides constraints that are  competitive with the (model-dependent) non-transient signature introduced in \cite{Stadnik:2020bfk}, which has, however, been derived for one specific sign of the coupling constant, see \cite{derevianko:arXiv2021,stadnik:arXiv2021}.
    Our clock approach based on the transient effect is therefore complementary to the torsion pendulum experiments based on the non-transient effect. This will be described in a future paper.     
\end{itemize}
Hence, this exploratory work, based on a simple MRA and on a simplified model for DM transient, paves the way for more refined and complex analysis and modeling. Nonetheless, its main result is a significant exclusion of the detection range of DM transients with size from 10$^5$ to 10$^9$ km for an encounter time $\mathcal{T}$ up to 2 months. 

\section{The GASTON model}\label{sec:model}

\subsection{Phenomenological model}

The coupling of ordinary matter or electromagnetism to a (light) scalar field $\varphi$ is motivated by certain effective theories that describe quantum gravity, such as string theory, see for example \cite{Uzan:2010pm,Damour:2010rp,Arvanitaki:2014faa,Hees:2018fpg}. In this study, we specifically opt for a quadratic interaction between the scalar field and the standard matter as in \cite{Stadnik:2015kia,Derevianko:2013oaa,Olive:2007aj}, in order to preserve the maximum of symmetries of the Lagrangian. From an effective point of view, these models predict a space variation of the fundamental parameters $\alpha$ and $\mu$ defined in (\ref{def:mu_ep}), in line with the phenomenological model of \cite{Damour:2010rp}, see also \cite{Bertrand:2020fho}. In the case a DM clump is formed from the condensation of the scalar field $\varphi(x)$, the effective fine structure constant and the fermion mass ``inside'' the clump are space-time dependent and read:
\begin{eqnarray}\label{eq:Effective-constant}
\alpha_{\text{eff}}(x) &=& \alpha_0\left( \frac{\Lambda_{\alpha}^2 \pm \hbar c\, V^2}{\Lambda_{\alpha}^2 \pm \hbar c\, \varphi^2(x)}\right), \\
m_{\text{eff}}(x) &=& m_\textrm{f}^0\, \left(\frac{\Lambda_f^2 \mp \hbar c\, \varphi^2(x)}{\Lambda_f^2 \mp \hbar c\, V^2}\right). \nonumber
\end{eqnarray}
As a reminder $\Lambda_{\alpha}$ and $\Lambda_f$ denote respectively the energy scale in the electromagnetic and fermionic sectors. The effective character of the fermion mass $m_{\text{eff}}(x)$ leads subsequently to a coordinate dependence of the proton-to-electron mass ratio $\mu$. The constants $\alpha_0$ and $m_\textrm{f}^0$ are respectively called the ``bare fine structure constant'' and the ``bare fermion mass''. It is supposed that the scalar field $\varphi(x)$ reaches its vacuum expectation value $V$ far from the clump (ffc). Therefore, this expression differs slightly from what can be found in the literature in order to ensure that:
\begin{equation}
    \lim_\textrm{ffc} \alpha_{\text{eff}}(x) = \alpha_0 \qquad \lim_\textrm{ffc} m_{\text{eff}}(x) = m_\textrm{f}^0 \, .
\end{equation}
Starting from an effective Lagrangian density, the physical consequences of such a change of notation are beyond the scope of this paper and will be addressed elsewhere. It is worthy to note that the expressions (\ref{eq:Effective-constant}) with the minus sign are only valid in the region of the parameter space where:
\begin{equation}\label{def:Regular-condition}
\hbar\, c\, \frac{V^2}{\Lambda^2_\alpha}<< 1 \, .
\end{equation}
In this section, we will define the coupling between the scalar field and the standard matter as: 
\begin{equation}\label{def:Gamma}
    \Gamma_X= \frac{\hbar\, c}{\Lambda^2_X}\, ,
\end{equation}
such that $\Gamma_X$ has dimension: $\textrm{m}^2\, \textrm{J}^{-2}$, namely the inverse dimension of $\varphi^2(x)$. Thus, the expression for the effective fundamental constants (\ref{eq:Effective-constant}) is identical for the SI and the natural unit system. Finally, the subscript $X$ refers to the fine structure constant, the proton mass or the mass of the electron according to the context. 

A possible overlap of atomic clocks by a transient DM clumps may induce an observable shift in atomic frequency $\omega_0$ due to the variation of the fundamental constants. Thus, under the condition (\ref{def:Regular-condition}), the fractional shift of a particular clock transition can be expressed as:
\begin{equation}\label{def:delta_omega}
\frac{\delta\omega}{\omega_0} = \pm \sum_X\kappa_X\Gamma_X\, \left[ V^2 - \varphi^2(\mathbf{r}, t) \right]\, ,
\end{equation}
where $\kappa_X$ are dimensionless sensitivity coefficients of effective changes in the constants $X$ for a particular clock transition and depends on experimental realization. For the satellite clocks of the Galileo constellation, the effective coupling constant $\Gamma_{\text{eff}}$ can be expressed in relativistic units, as determined in \cite{Guena:2012zz, Kozlov:2018}:
\begin{equation}\label{eq:gamma_eff}
\Gamma_{\text{eff}}(^1\textrm{H}) = 4\Gamma_{\alpha} + \Gamma_{\mu} ,
\end{equation}
where we have denoted:
\begin{equation}\label{def:Gamma_mu}
\Gamma_{\mu} \equiv \Gamma_{e/p} = 2\Gamma_{m_e} - \Gamma_{m_p} \, . 
\end{equation}

\subsection{Geometry of a transient encounter with the Earth}
\label{sec:geometry}

In the present work, we address the case in which the size of the DM clump exceeds the scale of the Galileo constellation.This allows us to employ the approximation of planar symmetry to describe the transient variation of the fundamental parameters. Moreover, we use the Domain Walls (DW), planar topological defects (see \cite{Vilenkin:2000jqa} for a review), as a toy model for DM clump formed through the condensation of a scalar field. Note that assessing the quality of this simplified network must be done on a case-by-case basis, according to the underlying fundamental model describing the variations of the fundamental parameters.

\begin{figure}[h!]
\centering
\includegraphics[width=0.45\textwidth]{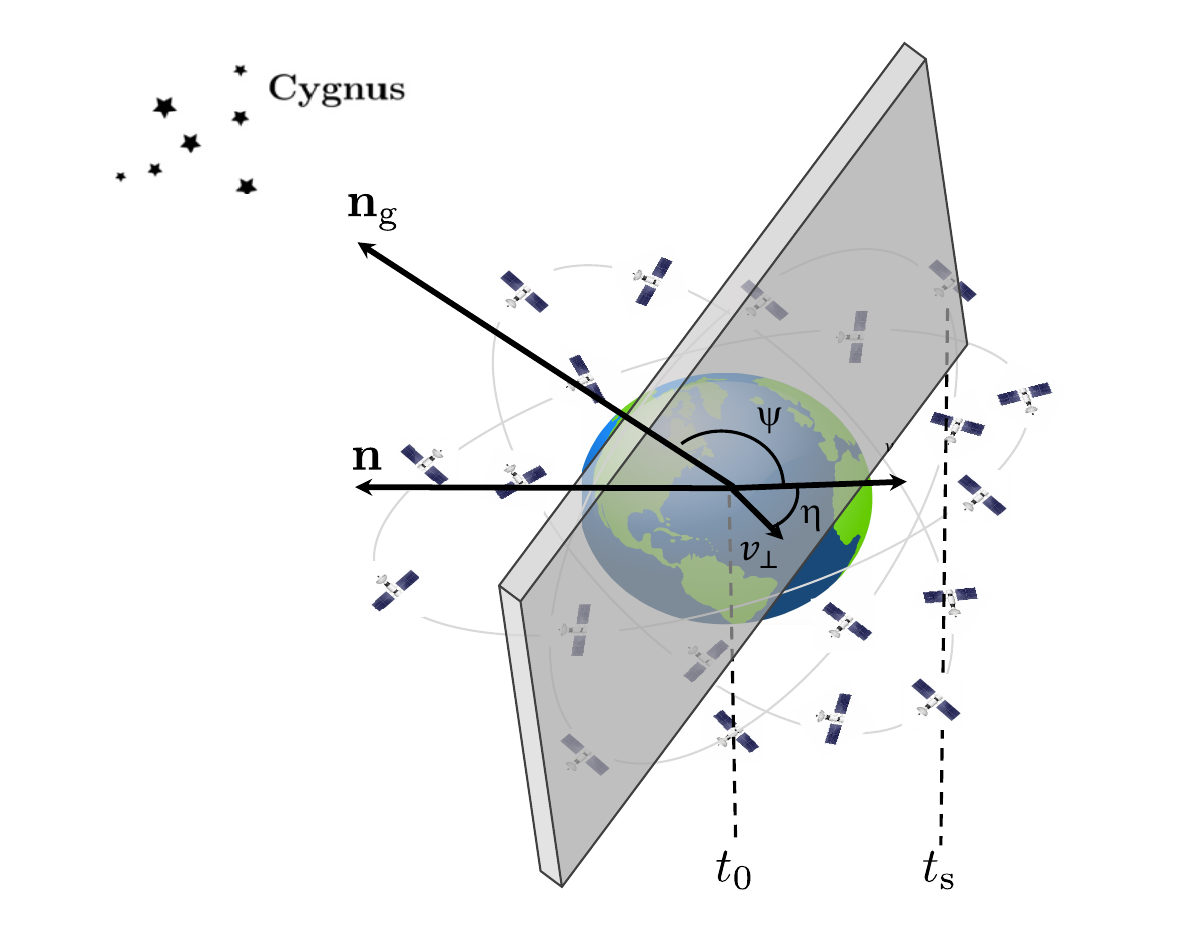}
\caption{Geometry of a transient passing at the center of the Earth at time $t_0$. The case illustrated here is the test bed model based on thin domain walls. The incident direction of the wall $\mathbf{n}$ is defined as $\mathbf{n} = -\mathbf{v}/|\mathbf{v}|$, with $\mathbf{v}$ being the relative velocity of the encounter. The velocity component perpendicular to the wall surface is denoted by $\mathbf{v}_{\perp}$. The thin DW crosses the satellite clock at $t_\textrm{s}$. }
\label{fig:DW_geometry}
\end{figure}

According to the standard halo model, the Milky Way rotates through the DM halo, and the Sun moves at a velocity $v_\textrm{c} \approx 220$ km/s towards the Cygnus constellation, see \cite{Bovy:2012tw}. This direction defines the most probable direction of incident DM clumps. In figure \ref{fig:DW_geometry}, inspired by \cite{Roberts:2018xqn}, the vector $\mathbf{n}_g$ points from the Earth center to this constellation. The transient DW crossing can be characterized by its relative velocity $\mathbf{v}$, its incident direction $\mathbf{n}$, its thickness $d$ and its time of encounter with the Earth $t_0$. We use the typical distance $L_\perp$ between pairs of Galileo satellites projected onto $\mathbf{n}$, defined later in section \ref{sec:Distance Threshold}, to make a distinction between ``thin'' and ``thick'' domain walls. Thin domain walls, for which $d < L_\perp$, represent a large transient DM object in which a strong variation of the scalar field occurs only close to the surface (``skin effect''). For the squared interaction (\ref{eq:Effective-constant}), thick domain walls, for which $d > L_\perp$, can be seen as an approximation for large DM clumps, regardless their shape, characterized by a smooth variation of the scalar field from the surface to the center. The Fig.\ref{fig:DW_geometry} illustrates the case of a thin domain wall. Nevertheless, the geometry of DM transient encountering the Earth is easily generalized to thick domain walls.  

In our model, we assume that a DW propagates with the velocity $\mathbf{v}$ and from the incident direction $\mathbf{n} = - \mathbf{v}/|\mathbf{v}|$, where the angle of the incidence $\psi$ is defined as $\mathbf{n}\cdot\mathbf{n}_\textrm{g} = -\cos\psi$. For the sake of convenience, we introduce the angle $\eta$, which defines the direction perpendicular to the wall surface $\mathbf{n}_{\perp}$. Then, the velocity component perpendicular to the wall surface, $v_{\perp}$, reads: 
\begin{equation}\label{def:v_perp}
    v_\perp = v\cos\eta \quad \textrm{and} \quad \mathbf{n}_{\perp} = - \mathbf{v}_{\perp}/|\mathbf{v}_{\perp}| \, .
\end{equation}
In this context, it is worth to mention that a DW  crossing the Galileo constellation with a velocity $\mathbf{v}$ cannot be distinguished from a DW with a  velocity vector $\mathbf{\tilde{v}}$ which is perpendicular to the wall surface $\mathbf{\tilde{v}}=\mathbf{v_\perp}$, provided \ref{def:v_perp} is satisfied. In fact, the velocity distribution can be introduced to generate a prior probability. Assuming the standard halo model, we consider again the quasi-Maxwellian model of \cite{Roberts:2017hla} describing the relative perpendicular velocity distribution of transient objects:
\begin{equation}\label{def:vperp_weight}
 f(v_\perp) =  C_\textrm{n}\, \frac{v_\perp}{v^3_\textrm{c}}\, \int_{v_\perp}^{\infty}  \left[\textrm{e}^{-\frac{(v-v_\textrm{c})^2}{v^2_\textrm{c}}} - \textrm{e}^{-\frac{(v+v_\textrm{c})^2}{v^2_\textrm{c}}} \right]\, \textrm{d}v \, ,
\end{equation}
where $v_\textrm{c}$ = 220 km/s is the Sun’s velocity in the halo frame while $C_\textrm{n}$ is a normalization constant such that
\begin{equation}
 C^{-1}_\textrm{n} =\int^\infty_0 f(v_\perp) \, \textrm{d}v_\perp    \, .
\end{equation}
We thus define the typical perpendicular velocity of the transient $v_\mathrm{gal}$ as the average of the perpendicular velocity:
\begin{equation}\label{def:vgal}
  v_\mathrm{gal} = \int_0^\infty\, v_\perp\, f(v_\perp)\, \textrm{d}v_\perp = 249 \textrm{ km/s}\, .
\end{equation}
In the model of \cite{Roberts:2017hla}, there is a strong cutoff for transient velocity at the galactic escape velocity (this argument was nevertheless discussed in \cite{Stadnik:2020bfk}). On the other hand, the system can never be static due to the spacecraft velocity, even though the transient perpendicular velocity turned out to be very low. The statistic Maxwellian distribution $f(v_\perp)$ for $v_\perp$ defined in (\ref{def:vperp_weight}) shows that 95\% of the events have a perpendicular relative velocity $v_\perp$ ranging from $v_{\perp\textrm{min}}$ = 46.22 km/s to $v_{\perp\textrm{max}}$ = 521.87 km/s. These bounds will therefore be used in this work. 
 
\subsection{Transient signal model}

\subsubsection{Network of transients}
The symmetries inherent in spherical transient DM clumps involve that the gradient of the scalar field is oriented along the radial direction. As a result, the large transient object can be assimilated to a Domain Wall (DW) with its quasi-planar surface perpendicular to the gradient of the scalar field. In this simplified framework, we recover the direction $\mathbf n_\perp$ and the associated radial coordinate $z$ of which the origin is at the core, i.e. the center of symmetry, of the transient.
We have chosen the profile of the scalar field $\varphi(z)$ in its own frame such that:
\begin{equation}\label{def:DW_phi-profile}
    \varphi(z;V,d) = V \tanh \left(\frac{z}{d}\right)\, ,
\end{equation}
where $V$ is the maximal amplitude of the scalar field outside the transient and $d$ is commonly referred to as the transient size, also related to the mass of the scalar field:
\begin{equation}
    d=\frac{\hbar}{c}\, \frac{2}{m_\varphi}\, .    
\end{equation}
In fact, the coupling of the field $\varphi(z)$ with ordinary matter varies significantly from $-2d$ to $2d$. Therefore, the zone of influence of the transient extends beyond $d$.

This choice is motivated by two reasons: 1) it corresponds to the simplest DW solution in a Klein-Gordon theory with a quartic self-interacting potential, which forms planar structures, as considered in \cite{Stadnik:2020bfk,Bertrand:2020fho} and 2) its square is supposed to mimic the squared radial profile of more complex spherical or cylindrical DM clumps, for a radius $d$ larger than $L_\perp$. As a reminder, the accuracy of this approximation should be assessed on a case-by-case basis.

Transient variations of fundamental parameter modeled by domain walls crossing the Earth possess an intrinsic energy cost, which is determined by the energy-stress tensor (\cite{Vilenkin:2000jqa}) that cannot exceed the local dark matter density $\rho_\textrm{DM}$. 
Hence, a limit on the energy density per domain wall $\sigma_\mathrm{DM}$ is reached in the extreme case where the energy density averaged over the domain wall network fulfills $\rho_\textrm{DM}$: 
\begin{equation}
  \sigma_\mathrm{DM}\leq\rho_\mathrm{DM}\, v_\mathrm{gal}\, \mathcal{T} \, ,
\end{equation}
where $\mathcal{T}$ is the mean duration between two consecutive encounters of transients with the Earth and, as a reminder, $v_\mathrm{gal}$ is their typical velocity. Therefore, the square of the amplitude $V$ of the domain wall is given by:
\begin{equation}\label{eq:DW-vev}
  V^2 = \frac{3}{4}\, v_\mathrm{gal}\, \mathcal{T}\, d\, \rho_\mathrm{DM} \, .
 \end{equation}
Notice that this result slightly differs from \cite{Derevianko:2013oaa} since we consider a smooth domain wall solution, more realistic  than a step function.  

\subsubsection{Maximum encounter time}

An important limit to the parameter space is related to the restrictions over the time of encounter $\mathcal{T}$: it must be statistically determined as being shorter than the duration of the observation campaign $\mathcal{T}_\textrm{camp}$, three months in our study. Indeed, during the observation campaign, we might expect theoretically the occurrence of $k=1$ event with $\mathcal{T} = \mathcal{T}_\textrm{camp}$, generating a frequency excursion of the clocks. Unfortunately, it may happen statistically that such an event does not happen during the observation campaign. In order to determine the maximum possible encounter time as a function of the campaign duration at a given level of statistical occurrence, we follow the hypothesis of \cite{Roberts:2017hla}. They assumed that the encounter period between the Earth and transient objects follows a Poisson distribution:
\begin{equation}
   P_k(\lambda) = \lambda^k\, \frac{\textrm{e}^{-\lambda}}{k!} \, .
\end{equation}
Let's suppose that, on average, we expect to see  $\lambda$ events during the campaign. This is an average, so that statistically we may observe fewer or more events than the $\lambda$ value. If we require theoretically that at least $k=1$ event occurs during this time with a statistical occurrence level of 95\%, we have:
\begin{equation}
 P_{k \ge 1}(\lambda) = 1 - P_0(\lambda) = 0.95\, ,
\end{equation}
which implies that $\lambda=3$ with the resulting limit $\mathcal{T}_\textrm{max}$ of the encounter time equal to:
\begin{equation}
 \mathcal{T}_\textrm{max} = \frac{\mathcal{T}_\textrm{camp}}{\lambda} = 1 \textrm{ month}\, .
\end{equation}

\subsubsection{Clock frequency bias}

Referring to the notations of section \ref{sec:geometry} and Fig. \ref{fig:DW_geometry}, the geocentric position  $\mathbf{X}(t)$ of any point located at the core of the DW is defined as:
\begin{equation}
\mathbf{X}(t) = (t - t_0) \, \mathbf{v}_{\perp} + \mathbf{y}\, ,
\end{equation}
where $\mathbf{y}$ is the projection of this position on the DW plane, satisfying $\mathbf{y}\cdot \mathbf{n}_{\perp} = 0$. Thereby, the distance between the DW and the satellite $a$ can be defined as:
\begin{eqnarray}
z_a(t; v_{\perp}, t_0) & = & (\mathbf{X}(t) - \mathbf{r}_a(t))\cdot \mathbf{n}_{\perp} \nonumber\\ 
 & = & - (t - t_0)\, v_{\perp} - \mathbf{r}_a(t)\cdot\mathbf{n}_{\perp},
\label{za}
\end{eqnarray}
where $\mathbf{r}_a(t)$ is the trajectory of the spacecraft $a$ in the ECI (Earth-centered Inertial) frame. 
In the case of an encounter between the Earth and a dark object, the observable is directly dependent on the interaction between ordinary matter and dark matter.
This interaction is characterized by a scalar field profile and amplitude $\Gamma_\textrm{eff}\, V^2$ derived from the Lagrangian density defined e.g. in \cite{Bertrand:2020fho,Stadnik:2020bfk}. 
In these studies, the hyperbolic tangent profile in (\ref{def:DW_phi-profile}) has been considered. 
We have here assumed that the amplitude $\Gamma_\textrm{eff}\, V^2(d;\mathcal{T},v_{\textrm{gal}})$ is a free parameter of the theory, with only the size parameter $d$ being specific to a given local event, the others depending directly on the global transient network, see (\ref{def:Gamma},\ref{eq:DW-vev}).
Indeed, the product $v_{\textrm{gal}}\, \mathcal{T}$ in (\ref{eq:DW-vev}) corresponds to the average distance between transients and is event-independent. On the contrary, the value of $v_\perp$ appearing in the definition (\ref{za}) is event-dependent.

Following the Eq. (\ref{def:delta_omega}), (\ref{eq:gamma_eff}) and (\Ref{def:DW_phi-profile}), the  clock phase bias induced by the DM clump in the clock $a$ can be computed at time $t$ as:
\begin{equation}
s_a(t) = \Gamma^a_{\text{eff}} \int_{-\infty}^t \left(\varphi^2(z_a(t';v_{\perp},t_0);V,d) - V^2\right)dt' \, .
\label{clock_bias}
\end{equation}
Unfortunately, the integral (\ref{clock_bias}) cannot be calculated analytically due to the dependence of $z_a(t)$ on the satellite Keplerian trajectory. Nevertheless, the first order difference for a time sample $\Delta T$ reads:
\begin{eqnarray} \label{first_order}
\Delta s_a(t) & = & s_a(t) - s_a(t-\Delta T) \\
& = & \Gamma_{\text{eff}}^a \int_{t - \Delta T}^t \Big(\varphi^2(z_a(t';v_{\perp},t_0);V,d) - V^2\Big)\, dt' \, , \nonumber
\end{eqnarray}
where the value $\Delta s_a(t)/\Delta T$ can be interpreted as a frequency, a term used later in the data analysis. If the spacecraft's trajectory is approximated by a linear motion,
\begin{equation}
    \mathbf{r}_a(t') = \mathbf{r}_a(t) + (t'-t)\, \mathbf{v}_a(t) \, ,
\end{equation}
where $\mathbf{v}_a(t)$ is the spacecraft velocity and $t' \in [t-\Delta T,t]$, the corresponding distance between the spacecraft and a transient object can be rewritten as:
\begin{eqnarray}
z_a(t'; v_{\perp},t_0) &=& -(t'-t_0)\, v_{\perp} - \mathbf{r}_a(t') \cdot \mathbf{n}_{\perp}
\\ \nonumber
&=& -(v_{\perp} + \mathbf{v}_a(t)\cdot \mathbf{n}_{\perp})\, t' + t_0\, v_{\perp} - \mathbf{r}_a(t) \cdot \mathbf{n}_{\perp} \nonumber\\ 
& & + t\, \mathbf{v}_a(t) \cdot \mathbf{n}_{\perp}\, . 
\end{eqnarray}    
Now, the distance $z_a$ is a linear function of time $t'$, which enables to integrate (\ref{first_order}) analytically. In the following, one can denote:
\begin{subequations}
\begin{eqnarray}
A_a(t) &=& -v_{\perp} - \mathbf{v}_a(t) \cdot \mathbf{n}_{\perp},
\label{eq:Aa}\\
B_a(t) &=& v_{\perp}\, t_0 - \mathbf{r}_a(t) \cdot \mathbf{n}_{\perp} + t\, \mathbf{v}_a(t) \cdot \mathbf{n}_{\perp},
\label{eq:Ba}
\end{eqnarray}
\end{subequations}
such that:
\begin{equation}\label{eq:zat}
    z_a(t';v_\perp,t_0) = A_a(t)\, t' + B_a(t)\, .
\end{equation}
%
Using these notations and Eq.~(\ref{eq:zat}), the result of the integration of (\ref{first_order}) with the scalar field profile introduced in (\ref{def:DW_phi-profile}) can be written in a streamline notation:
\begin{equation}\label{DeltaS}
\Delta s_a(t) = \frac{\Gamma_{\text{eff}}^a\, V^2\, d}{A_a(t)}\left[\tanh\left(\frac{z_a(t-\Delta T)}{d}\right) - \tanh\left(\frac{z_a(t)}{d}\right)\right] \, .   
\end{equation}
%

\subsection{Simulated signals for independent pairs of clocks} \label{sec:pairs}
\begin{figure}[h!]
\centering
\includegraphics[width=0.42\textwidth]{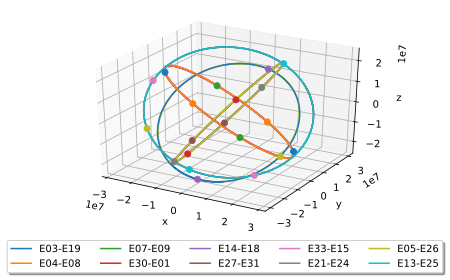}
\caption{Galileo orbits with points referring to the position of the different satellites at some random time.
}
\label{av_distance}
\end{figure}

%
The Galileo constellation consists of 26 satellites distributed in three quasi-circular orbits, each with a radius of $d = 3\times 10^4$ km and  inclined at an angle of 56 degrees to the equator. %
Their revolution period is about 14 hours. Our study makes use of the high stability
of the onboard H-masers.
It will be explained in subsection \ref{sec:MRA_Large-DW} that the further apart the 2 satellites are, the stronger the signal. For this reason, we propose here simulations for 10 pairs of satellites, which have been chosen to maximize the between-satellite distances while ensuring that each satellite appears only once. These pairs are represented in Fig.\ref{av_distance}. 

\begin{figure*}[ht!]
\centering
\includegraphics[width=0.6\textwidth]{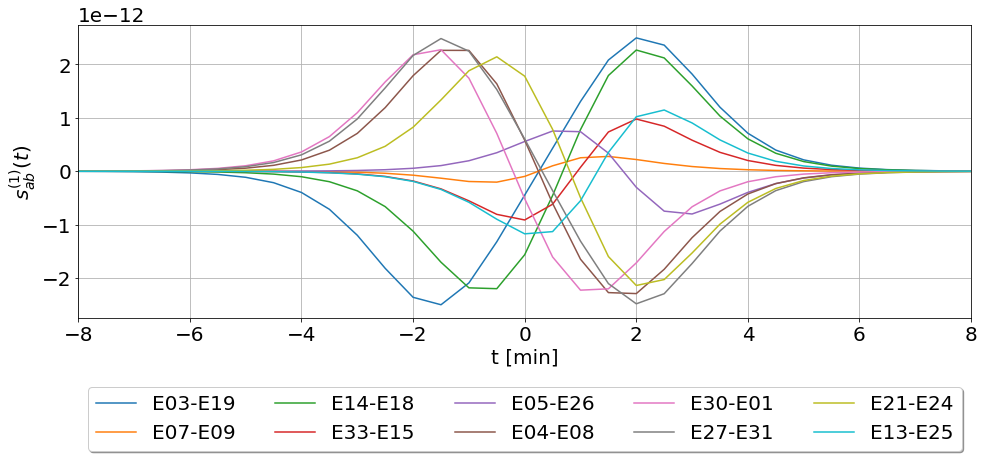}
\caption{Simulated relative frequency bias  $s_{ab}^{(1)}(t)$ for $10$ pairs of Galileo clocks for a thin DW with $v_{\perp} = 220 \text{km/s}$, $t_0 = 0$ min, $\Lambda_\alpha = 6\times 10^7$ TeV, $\mathbf{n}_{\perp} = (1, 0,0)$, and a thickness $d = 2\cdot 10^4 \text{km}$.}
\label{fig:sim-sig_thin-DW}
\end{figure*}

\begin{figure*}[ht!]
\centering
\includegraphics[width=0.6\textwidth]{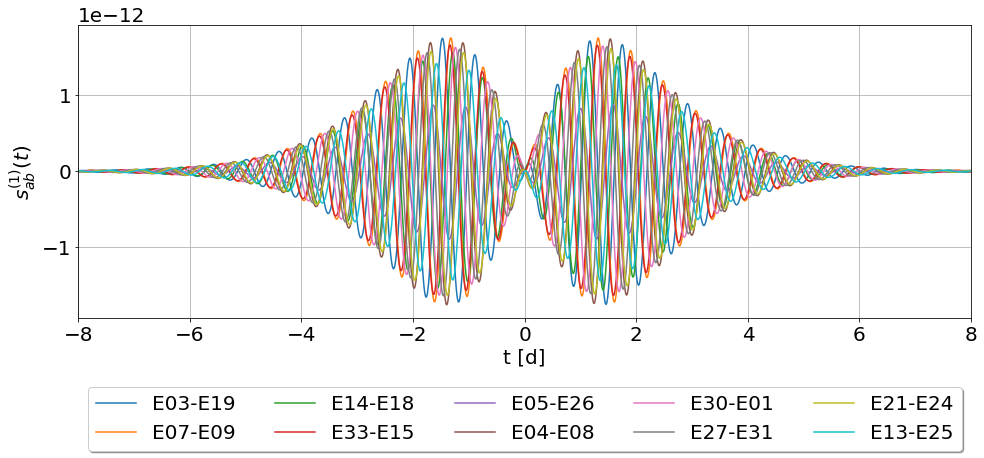}
\caption{Simulated relative frequency bias  $s_{ab}^{(1)}(t)$ for $10$ pairs of Galileo clocks for a large transient with $v_{\perp} = 220 \text{km/s}$, $t_0 = 0$ d, $\Lambda_\alpha = 10^8$ TeV, $\mathbf{n}_{\perp} = (1,0,0)$, and $d = 10^7 \text{km}$.}
\label{fig:sim-sig_thick-DW}
\end{figure*}

Considering a particular pair of clocks onboard Galileo satellites, denoted by $a$ and $b$, their relative phase and frequency biases read:
\begin{subequations}\label{eq:signal_ds}
\begin{eqnarray}
s_{ab}^{(0)}(t) &=& s_{a}(t) - s_{b}(t),
\\
s_{ab}^{(1)}(t) &=& \frac{\Delta s_{a}(t)}{\Delta T} - \frac{\Delta s_{b}(t)}{\Delta T}\, . \label{eq:dsa}
\end{eqnarray}
\end{subequations}
The frequency bias $s_{ab}^{(1)}(t)$ induced by a thin DW with a thickness $d = 2\times 10^4$ km is shown in fig. \ref{fig:sim-sig_thin-DW} while the case of a large transient with  $d = 10^7$ km is displayed in fig. \ref{fig:sim-sig_thick-DW}. For both simulations, we considered a transient with a perpendicular component of velocity $v_{\perp} = 220$ km/s and a normal direction $\mathbf{n}_{\perp} = (1, 0, 0)$. We have considered in these simulations the model-dependent parameter $\mathcal{T} = \mathcal{T}_\textrm{max}$ = 1 month.

For a thin and fast DW, i.e. $v_{\perp} >> \mathbf{v}_a\cdot \mathbf{n}_{\perp}$, the spacecrafts positions and velocities do not contribute significantly to the coefficients $A_a(t)$ and $B_a(t)$ in Eq. (\ref{eq:Aa}) and (\ref{eq:Ba}). The typical signature at the clock satellite level is either a frequency spike or an over-damped frequency periodic oscillation.
For a large transient, fig.\ref{fig:sim-sig_thick-DW} shows that the Galileo satellite motion is no longer negligible: the typical signature is a periodic variation of the frequency  bias $ s^{(1)}_{ab}$ at the Galileo satellite orbital period. This variation is modulated by an envelope depending on the DM transient size. This case is addressed and modeled for the first time.
This periodicity arises from the fact that the gradient of the scalar field in the $\mathbf{n}_{\perp}$-direction induces a phase difference between the clocks onboard the satellites of the pair  ($a,b$). For example, if the gradient is negative, any clock $a$ ahead of clock $b$ in the $\mathbf{n}_{\perp}$ direction will be delayed relative to clock $b$. After half an orbital period, the same clock $a$ will be behind clock $b$, causing it to be ahead in phase with the latter. This pattern repeats  for several orbital periods, generating a sinusoidal signature.

\section{The maximum reach analysis}\label{sec:Max-reach_analysis}

A simple detection of a maximum frequency spike provides an initial strong exclusion area in the parameter space. This method is a modified version of the ``Maximum Reach Analysis'' (MRA) which enables a rapid estimation of the sensitivity of a parameter without the need for an exhaustive statistical analysis. We therefore determine the maximum theoretical value that a specified parameter can assume without causing a frequency spike larger than what is measured in the dataset.

\subsection{Methodology}

Our MRA takes into account the maximum frequency values in the $s^{(1)}_{ab}(t)$ dataset and compares them with the worst-case value predicted by theoretical modeling. This modeling depends on a set of parameters ($t_0$, $\bm v_\perp$, $d$), as well as $V(d,\mathcal{T})$, which will be chosen to yield the most conservative constraint on the coupling parameter $\Gamma_\mathrm{eff}$. 
%
This method drastically reduces the computational demand and allows the exploration of a very large region of the parameter space.
The drawback is that it takes no account of the prior distribution of parameters, and furthermore relies on a single measure: the ``maximum reached''. It therefore does not exploit the large number of data points which are available for conducting a correlation search. Nevertheless, this method can already be used to derive preliminary constraints on the parameter space. Subsequently, it excludes event configurations in which  the effort to detect transients is not justified, given their unlikeliness.

Let's provide a more detailed description of the MRA  methodology. For each clock $a$, we consider the frequency difference with all the other clocks $b$ and look for the maximum frequency peak in the $s^{(1)}_{ab}(t)$ dataset, with no regard for past frequency evolution.
More precisely, for each clock $a$, we find $s^{(1)}_{a,\mathrm{max}}$ defined as:
\begin{equation}\label{def:DeltaS_max}
    s^{(1)}_{a,\mathrm{max}} = \mathrm{max}_{b\neq a}\quad \mathrm{max}_{t_i}\  s^{(1)}_{ab}(t_i) \, ,
\end{equation}
in the dataset where each epoch is denoted by $t_i$. 

On the other hand, Eqs.~(\ref{DeltaS}) and (\ref{eq:dsa}) allow writing the theoretical modeling as:
\begin{equation}\label{def:s-tilde}
    \Gamma_\mathrm{eff}\, V^2(d,\mathcal T)\, \tilde s_{ab}(t;\bm v_\perp, t_0, d;\bm r_a,\bm r_b) = s^{(1)}_{ab}(t)\, ,
\end{equation}
where $\tilde s_{ab}(t,\xi,\bm r)$ is a pattern, depending on a set of parameters $\xi$ characterising the transient encounter:
\begin{equation}\label{def:xi}
    \xi \equiv (\bm v_\perp, t_0, d)\, ,
\end{equation}
while $V^2(d,\mathcal T)$ is given by Eq. (\ref{eq:DW-vev}).
The idea of the MRA is to find the function $s_*(d)$, that saturates the inequality:
\begin{equation} \label{def:s*}
   s_*(d)\leq  \max_{t,a,b}\, \tilde s_{ab}(t;\bm v_\perp, t_0, d;\bm r_a,\bm r_b)\, ,
\end{equation} 
for all possible satellite configurations $(\bm r_a,\bm r_b)$.
This means that $s_*(d)$ acts as a lower bound for a set of theoretical patterns $\tilde s_{ab}(t,\xi,\bm r)$ for any $t$ and $t_0$ included in our dataset time baseline, for the ``worst'' satellite configuration that minimizes $\tilde s_{ab}$, and for 95 \% of the values for the parameters $\mathcal{T}$ and $\bm v_\perp$.

Then, the actual constraint on $\Gamma_\mathrm{eff}$ as a function of $d$ is given by:
\begin{equation}\label{def:constraints}
   \Gamma_\mathrm{eff}\leq \frac{\mathrm {min}_a\ s_{a,\mathrm{max}}^{(1)}}{V^2(d,\mathcal T)\, s_*(d)}\, . 
\end{equation}
The challenge lies in finding a reliable value for $s_*$ that reaches the lower limit on the signatures $\tilde s_{ab}(t,\xi,\bm r)$. This is done in two steps: 
\begin{enumerate}
    \item Examine the geometry of the constellation to define a distance threshold, namely a lower bound on the maximal projected distance between satellites.
    \item Derive an analytical expression for $s_*(d)$ by determining the values of $z_a$ that appear in the argument of the hyperbolic tangent (Eq.~(\ref{DeltaS}), in the cases of  thin DW and of  large transients.
\end{enumerate}

\subsection{Distance threshold}\label{sec:Distance Threshold}

Let's consider two satellites $a$ and $b$. The relative frequency variation induced by the transient on their clock differences depends on the distance between the two satellites, projected onto the normal $\mathbf{n}_\perp$ to the transient surface. The projection of the satellite $b$ position $\mathbf{r}_b(t)$ may indeed be written as:
\begin{equation}
 \mathbf{r}_b(t) \cdot \mathbf{n}_{\perp} = L_{\perp{ab}}(t) + \mathbf{r}_a(t) \cdot \mathbf{n}_{\perp}\, ,
\end{equation}
in the expression for the frequency spike (\ref{eq:dsa}) along with (\ref{DeltaS}). We can thus rewrite the set of physical patterns defined in (\ref{def:s-tilde}) as a function of the projected distance:
\begin{equation}
 \tilde s_{ab}(t;\bm v_\perp, t_0, d;\bm r_a,\bm r_b) \to \tilde s_{ab}(t;\bm v_\perp, t_0, d;\bm r_a,L_{\perp\rm{ab}})\, .
\end{equation}
In the MRA framework, the encounter time $t_0$ is irrelevant: the largest frequency spike can occur at any time during the 3-month campaign. Moreover, our MRA does not weight the probability distribution of the angle of incidence of the transient but instead considers 1) the largest effect 2) in the most unfavorable configuration. 
1) For a large transient, the largest phase delay between two clocks is obtained when their projected distance $L_{\perp{ab}}$ is maximum. 
2) We define a lower bound for the maximum projected distance between the satellites $a$ and all other satellites, denoted $L_{\perp}$. This ensures that, at any time, there is at least one satellite $b$ for which the projected distance between $a$ and $b$ exceeds $L_{\perp}$.

\begin{figure}[h!]
\centering 
\includegraphics[width=0.6\linewidth]{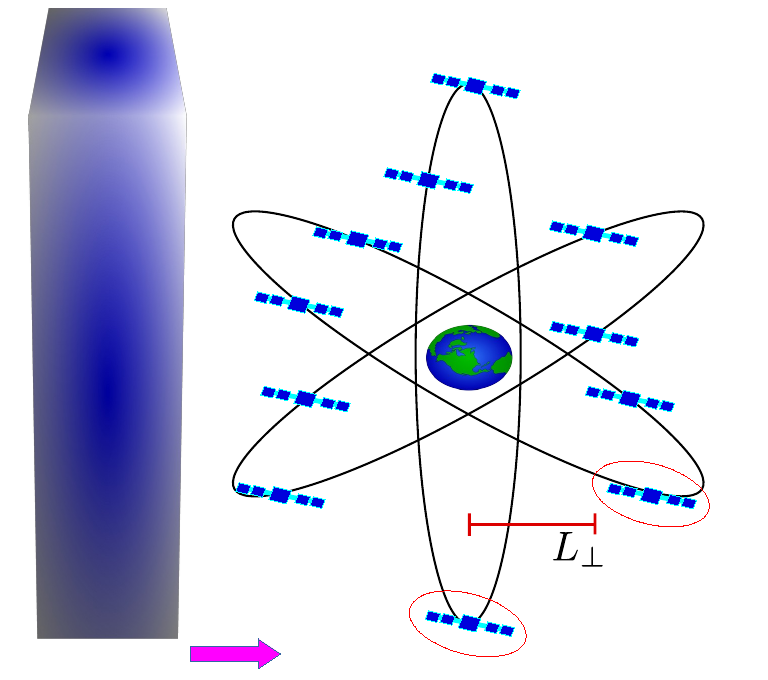}
\caption{\small{Illustration of the distance threshold $L_{\perp}$ when the transient surface is parallel to one of the orbital plane.} \label{fig:orbital-planes}}
\end{figure}

To this end, we determine, for any Galileo satellite $a$, the distance of the farthest satellite when we are in the worst configuration, i.e. when this distance is the shortest. 
This worst-case configuration occurs when the transient surface is almost parallel to one of the three Galileo's orbital planes, see fig.\ref{fig:orbital-planes}. In this case, all pairs of Galileo clocks belonging to the same orbital plane exhibit no effect. In addition, the three orbital planes in the Galileo constellation present an inclination angle of 56$^\circ$ to the equatorial plane. So, for any satellite close to the equatorial plane at the time of an event, the projected distance to the other satellites remains limited. Subsequently, the distance threshold projected onto $n_\perp$, reads:
\begin{equation}\label{def:Lar}
 L_{\perp} = R_{\textrm{sat}}\cos(30^\circ) \sqrt{1-\frac{1}{4}\, \sin^2(68^\circ)\, [\sin(60^\circ) + 1]^2} ,
\end{equation}
where $R_{\textrm{sat}}$ is the orbital radius. The angle of 68$^\circ$ is the double of the complementary angle of the inclination angle (56$^\circ$). The angle of 60$^\circ$ comes from the azimuth angle between the orbital planes (120$^\circ$). The angle of 30$^\circ$ comes from the fact that we have considered 6 satellites per orbital plane. It's a conservative hypothesis since we did not take into account the eccentric Galileo satellites E14 and E18 in the modeling while they are included in the data series. From equation (\ref{def:Lar}), we find $L_\perp = 13000$ km.

\subsection{Signal modeling}


%

The specific hyperbolic tangent model introduced in (\ref{def:DW_phi-profile})
allows writing the expression (\ref{DeltaS}) in the simple analytical form:
\begin{equation}\label{eq:DeltaS_Max-frequency}
    s_{ab}^{(1)} = \frac{\Gamma_\mathrm{eff}\, d\, V}{\Delta T}\, \left[\frac{\Delta\varphi_{a}}{A_a} - \frac{\Delta\varphi_{b}}{A_b}\right] \, .
\end{equation}
In the above expression (\ref{eq:DeltaS_Max-frequency}), the difference $\Delta \varphi_{a}$ relative to the initial position of the clock $a$ and the distance covered by the transient reads:
  \begin{equation}
   \Delta \varphi_{a} = \varphi\left(z_a(t-\Delta T)\right)-\varphi\left(z_a(t)\right)\, .
  \end{equation}
%
As a reminder, the Maxwellian distribution $f(v_\perp)$ for $v_\perp$ defined in (\ref{def:vperp_weight}) shows that 95\% of the events have a perpendicular relative velocity $v_\perp$ ranging from $v_{\perp\textrm{min}}$ = 46.22 km/s to $v_{\perp\textrm{max}}$ = 521.87 km/s.
%
Introducing $v_*$ such that $v_* = v_\perp+\bm v_a\bm n_\perp$,
the goal of the MRA is to find the lower bound $s_*(d)$ (\ref{def:s*}) for the set of theoretical patterns $\tilde s_{ab}(t,\xi,\bm r)$:
\begin{align}\label{eq:tilde_sab}
    \frac{v_*\, \Delta T}{d}\, s_*(d) 
    \leq \max_{t,a,b}\, & \left[ \tanh\left(\frac{z_a+v_*\, \Delta T}{d}\right)-\tanh\left(\frac{z_a}{d}\right) \right. \nonumber\\
    & \left. - \tanh\left(\frac{z_b+ v_*\, \Delta T}{d}\right)+\tanh\left(\frac{z_b}{d}\right) \right]\, , 
\end{align}
by fixing $v_*$. The question is: how to determine the value of $v_*$ either from $v_{\perp\textrm{min}}$ or from $v_{\perp\textrm{max}}$.
%
%
In the same way, the ``$\max_{t,a,b}$'' applies to all possible satellite configurations given by $(z_a,z_b)$.
As a consequence, the parameters describing the projected satellite positions $z_a$ and $z_b$ will be chosen based on the least favorable satellite configuration, i.e. the configuration that minimizes the maximum of the right hand side of the inequality  Eq.~(\ref{eq:tilde_sab}). This configuration is defined differently whether thin domain walls or large transients are considered. They will be distinguished using the $L_{\perp}$ scale. 

\subsubsection{Large transients}\label{sec:MRA_Large-DW}

\begin{figure}[h!]
\centering 
\includegraphics[width=0.75\linewidth]{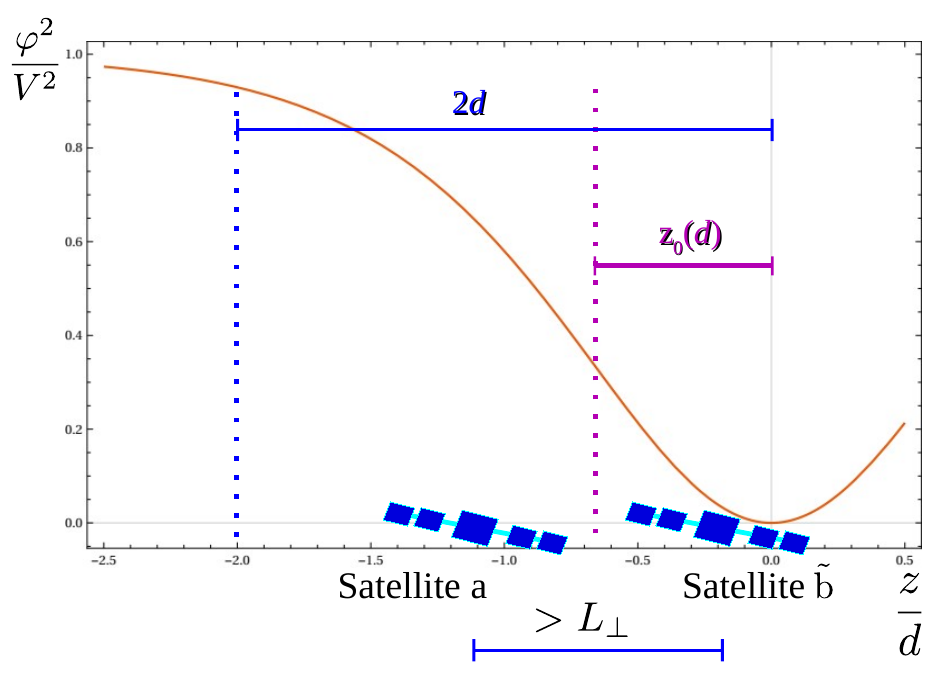}
\caption{\small{Optimal configuration between two clocks $a$ and $\tilde{b}$ for a maximum frequency spike $s^{(1)}_{ab}$ during the transit of a dark object larger than the typical Galileo constellation scale: $d > 9900$ km.}\label{fig:Max-jump_Large-transient}}
\end{figure}

In the case of thick domain walls, which can potentially represent some large transients of any shape, we have already shown that the effect induced by the transient on clocks is proportional to the gradient of the scalar field.
In case of event, we only consider the moment when the relative frequency $s_{ab}^{(1)}$ between a given satellite $a$ and all other Galileo satellites is maximum, ensuring that at least one satellite, denoted $\tilde{b}$, is located at a projected distance larger than $L_{\perp}$. 
For that purpose, in the transient reference frame, we define the coordinate $z_\textrm{0}$ as the point where the space variation of the effective fundamental constants defined in (\ref{eq:Effective-constant}) is maximum. Actually, it corresponds to the extremum of the first order derivative of the scalar field coupling profile. Then, considering the hyperbolic profile (\ref{def:DW_phi-profile}), such a condition arises for $z_0 = \pm0.658\, d$. In this section, we will choose arbitrarily the negative value $z_0(d) = -0.658\, d$. Hence, in equation (\ref{eq:tilde_sab}), we fix the value of $z_{a}(t)$ and $z_{b}(t)$ according to the optimal configuration where the two satellites $a$ and $\tilde{b}$ are located on either side of the coordinate $z_0$, as illustrated in fig.\ref{fig:Max-jump_Large-transient}.  
According to all the above considerations, $z_{a}(t)$ and $z_{b}(t)$ are thus time-independent and read:
\begin{eqnarray}\label{def:MaxJump_B_LT}
 z_{a} & = & z_0 - \frac{L_{\perp}}{2}\, , \\
 z_{b} & = & z_0 + \frac{L_{\perp}}{2}\, .  \nonumber
\end{eqnarray}
The validity of the definition of $z_{b}$ directly suggests the transition between large transient and thin domain wall regimes, since it requires that $-2 z_0(d) > L_{\perp}$, so $d > 9900$ km. 
 Furthermore, the function (\ref{eq:tilde_sab}) is monotonically decreasing on the variable $v_*$ for $d > 9900$ km. Consequently, the value of $v_*$ will be taken at the 95\% upper limit of the $v_\perp$ distribution, $v_{\perp\textrm{max}}$, in addition to the typical velocity for a Galileo satellite (i.e. 3.7 km/s).

For the very large values of $d$, a Taylor expansion of (\ref{eq:tilde_sab}), with the satellite position (\ref{def:MaxJump_B_LT}), can be used to show that the value of $s_*$ defined in (\ref{def:s*}) is independent of $v_\perp$:
\begin{equation}
   s_*(d) \approx \cosh^{-2}\left(\frac{z_{a}}{d} \right) - \cosh^{-2}\left(\frac{z_{b}}{d} \right) \, .
\end{equation}
This expansion requires $d > 33900 \mathrm{km}$ through the condition:
\begin{equation}
    \textrm{Min}\left(|z_a|,|z_b|\right) \gg v_*\, \Delta T \, .
\end{equation}


\subsubsection{Thin domain walls}
\begin{figure}[h!]
\centering 
\includegraphics[width=0.75\linewidth]{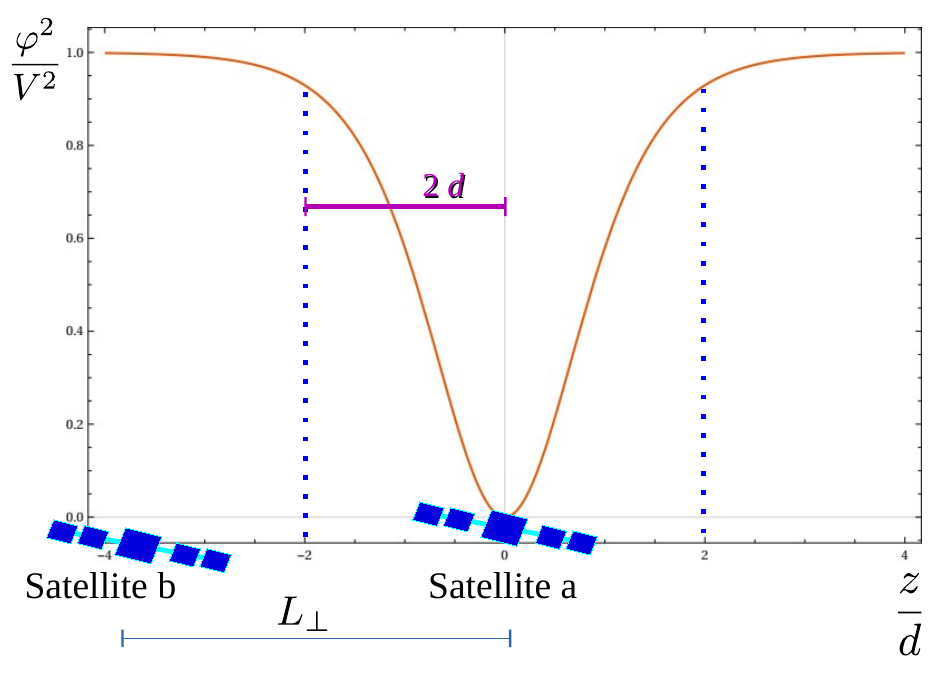}
\caption{\small{Configuration between two clocks $a$ and $b$ in case of thin domain walls characterized by $d < 6500$ km.}\label{fig:Max-jump_thin-wall}}
\end{figure}

A domain wall is considered as "thin" when its thickness $d$ is smaller than the typical distance between Galileo satellites. Hence, the configuration optimizing the transient effect occurs when one of the two satellites of the pair ($a$,$b$) is located at the core of the domain wall. This happens twice during the transit, as illustrated in fig.\ref{fig:Max-jump_thin-wall}. According to these assumptions:
%
\begin{eqnarray} \label{def:MaxJump_Thin-DW}
 z_{a} & = & 0 \, , \\
 z_{b} & = & \pm L_{{ab}\perp} \, , \nonumber
\end{eqnarray}
for each satellites pair ($a$,$b$) with $L_{{ab}\perp} > 2\, d$. According to the definition of the distance threshold $L_\perp$, for a given satellite $a$ it is always possible to find at least one satellite $b$ that fulfills this condition. Therefore, the condition on $d$ for considering a domain wall as thin is the following:
\begin{equation}
    d < \frac{L_\perp}{2} = 6500 \textrm{ km}\, .
\end{equation}
Hence, using (\ref{eq:tilde_sab}) and the satellite position (\ref{def:MaxJump_Thin-DW}), the lower bound for thin domain walls reads at a 95\% of occurrence:
\begin{eqnarray}
 \frac{v_*\, \Delta T}{d}\, s_*(d) & = & \tanh\left( \frac{v_*\, \Delta T}{d} \right) - \tanh \left(\frac{v_*\, \Delta T \pm L_{\perp}}{d}\right) \nonumber\\
 && \pm \tanh \left(\frac{L_{\perp}}{d}\right) \, .
\end{eqnarray}
As before, $v_*$ was fixed by taking the 95\% upper limit of the $v_\perp$ distribution, $v_{\perp\textrm{max}}$.

%

\subsubsection{Blind area for $6500 < d < 9900$ km.}

Intermediate values of $d$ with $6500 < d < 9900$ km correspond to the transition between thin domain walls and large transients. 
In this case, it is difficult to fix analytically and unequivocally the position of the satellites and their threshold distance, as in Eq.~(\ref{def:MaxJump_B_LT}) and (\ref{def:MaxJump_Thin-DW}). If one of the latter conditions is used, we recover the phenomenon of clock degeneracy described in \cite{Roberts:2017hla}. The function (\ref{eq:tilde_sab}) is no longer monotonic in $v_*$ and can even be zero for some combinations of $d$ and $v_*$.
%
%
%
%
Therefore, our MRA has a ``blind'' area in the parameter space where it is not possible to analytically define a lower bound that saturates the inequality (\ref{eq:tilde_sab}) by fixing the value of $v_*$ and choosing a simple satellite configuration with $(z_a,z_b)$. 
%

\subsection{Data mining}
\label{sec:data-mining}

The clock data have been computed by  the ESOC analysis center of the International GNSS Service (IGS), from the measurements collected in a network of ground stations. The satellite clock data are reported with a sampling rate $\Delta T$ of 30s with respect to a stable ground reference. Clock phase biases $s_{a}^{(0)}(t)$ for each satellite $a$ have been computed over a 3-month period from 2021 January 1st to April 1st.
In order to remove any possible disturbances from the ground reference clocks of the ESOC products, we work with the frequency differences between satellite clocks $a$ and $b$ (\ref{eq:dsa}).  
The data have been analysed using moving windows of 14 hours, i.e. the orbital period of the Galileo constellation. In each 14h window and for each satellite pair, we removed a linear frequency frequency drift to keep only the frequency variations. As explained above, the signature of large transients in  $s^{(0)}_{ab}(t)$ and $s^{(1)}_{ab}(t)$ is a 14h periodic signal. The choice of a 14h window ensures that  possible periodic signals are not affected when removing the linear frequency drift. However, this 14h window imposes using data from successive days in a same window. This induces frequency outliers at the day boundaries,  due to the phase jumps inherent to the processing of the clock products in daily batches. Finally, we took the maximum frequency spike among all 14h windows. 
 
 

%
In addition to the day-boundary frequency outliers mentioned here above, some others appear at other epochs. In order to check if their origin is from a real signal or an artefact to the clock products used in our analysis, we compared our data with those obtained clock products from another analysis center (AC) of the IGS. The frequency spike was removed only if it was confirmed in the other AC clock solutions.   

\subsection{Placing limits}

\subsubsection{Experimental data series}

\begin{figure}[h!]
\centering 
\includegraphics[width=1.01\linewidth]{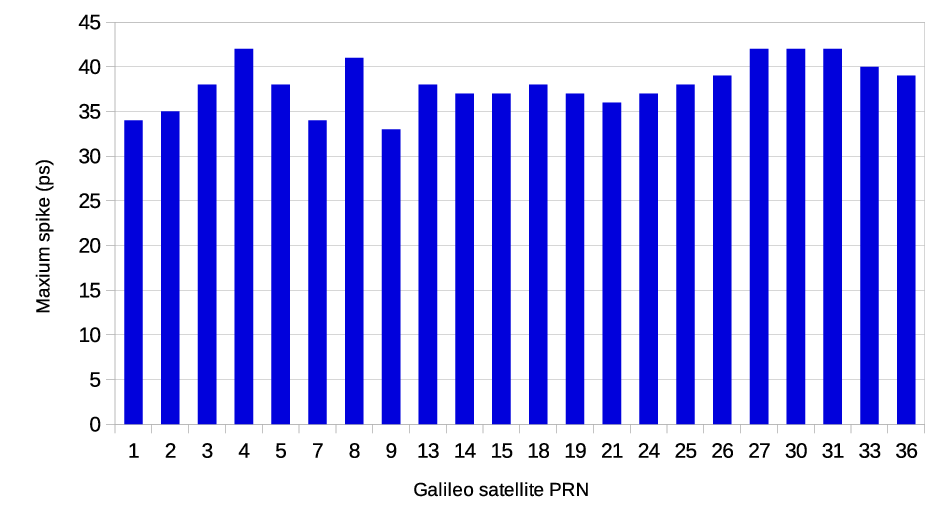}
\caption{\small{Maximum value $\Delta s^a_{\mathrm{max}}$ obtained from Galileo satellite $a$ over the 3-month campaign.}\label{fig:Maximum-Jump_sat-pairs}}
\end{figure}

As a reminder, for each Galileo spacecraft $a$, we considered the frequency differences $s^{(1)}_{ab}(t)$ with all the other satellites $b$ and retained the maximum frequency spike $s^{(1)}_{{a},\mathrm{max}}$ over the whole 3-month campaign. Our experimental results for $s^{(1)}_{{a},\mathrm{max}}$ are displayed in Fig.~\ref{fig:Maximum-Jump_sat-pairs}. 
In order to keep streamline notations, we have introduced in this section the experimental value $\Delta s_{{a},\mathrm{max}}$ defined by:
\begin{equation}
 \Delta s^a_{\mathrm{max}} = \Delta T\, s^{(1)}_{{a},\mathrm{max}}\, .
\end{equation}

\subsubsection{Experimental frequency threshold}\label{sec:Freq_threshold}

\begin{figure}[h!]
\centering 
\includegraphics[width=1.0\linewidth]{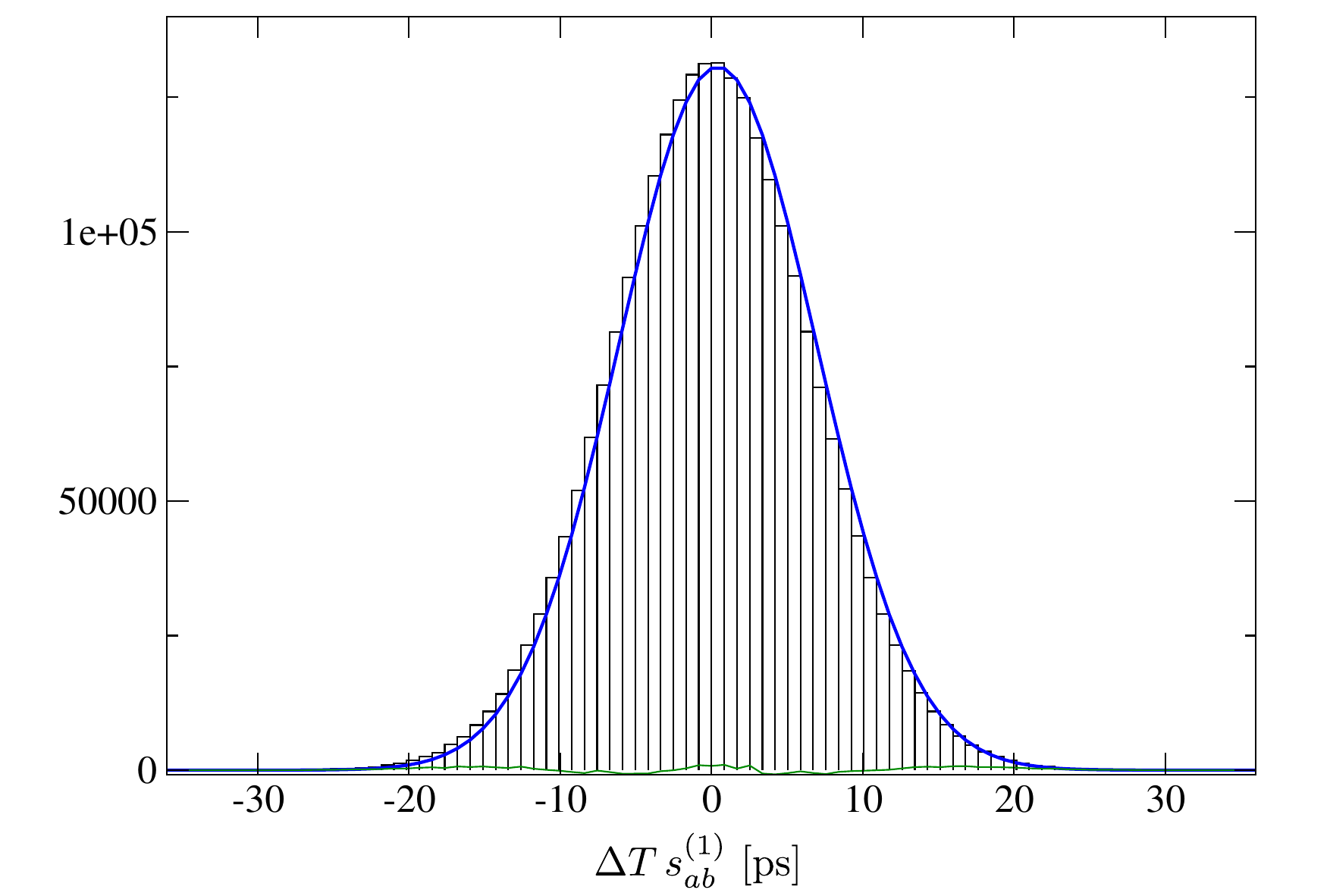}
\caption{\small{Histogram of the data series $\Delta T\, s^{(1)}_{ab}(t)$ (defined by Eq.~\ref{eq:dsa}) for an set of 10 independent clock pairs $ab$ defined in section \ref{sec:pairs}. The blue curve is the fitted Gaussian used to determine $\sigma_c$. The green curve is the plot of residuals.}\label{fig:Gaussian-histogram}}
\end{figure}

In the event of a transit by a dark object, we assume that all the satellite pairs with a (projected) distance larger than the threshold $L_{\perp}$ must have experienced a detectable frequency excursion.
Therefore, we consider as a ``no detection level'' associated to $L_{\perp}$ the minimum value of  the set $\Delta s^a_{\mathrm{max}}$ obtained, which is 33 ps as seen in fig.\ref{fig:Maximum-Jump_sat-pairs}. 
%
%
However, it can happen that one clock of the pairs has experienced a statistical fluctuation during the epoch where the maximum $\Delta s^a_{\mathrm{max}}$ arises. In order to consider this possibility, we determine the typical uncertainty $\sigma_c$ of the frequency comparisons  by fitting a Gaussian profile (fig.\ref{fig:Gaussian-histogram}), which leads to $\sigma_c=6.5$ ps. This Gaussian fit was generated from the frequency differences of the ten independent pairs of clocks used in section \ref{sec:pairs}, which ensures the use of uncorrelated data. Hence, the experimental frequency threshold value:
\begin{equation}
 \mathrm {min}_a\ \Delta s^a_{\mathrm{max}} = 33 + 2\, \sigma_\textrm{c}\quad [\textrm{ps}] \, ,
\end{equation}
fulfills the constraints on $\Gamma_\textrm{eff}$ in Eq. (\ref{def:constraints}):  
\begin{equation}\label{def:freq-thres_LT}
 \Gamma_\textrm{eff}\, V^2\, \Delta T\, s_* \leq 46\, \textrm{ps}\, .
\end{equation}

\subsubsection{Exclusion area in the parameter space}

\begin{figure}[h!]
\centering 
\includegraphics[width=0.85\linewidth]{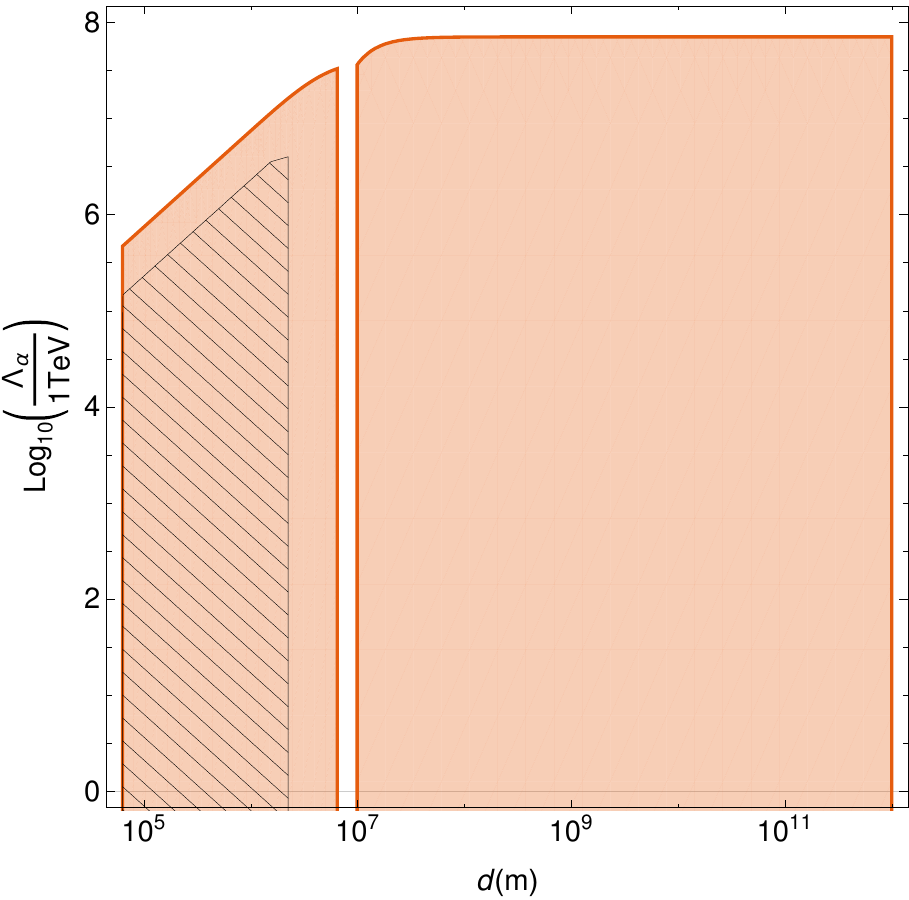}
\caption{\small{Graph of constraints on the parameter space with the energy scale $\Lambda_\alpha$ versus the size of the transient $d$, for $\mathcal{T}_\textrm{max}$ = 1 month and assuming $\Gamma_\alpha \gg \Gamma_\mu$ in (\ref{eq:gamma_eff}). The hatched region represents the constraints previously achieved in \cite{Roberts:2017hla}. 
The exclusion area in orange results from the MRA introduced in this work.}\label{fig:Max-jump_Constraints}}
\end{figure}

\begin{figure}[h!]
\centering 
\includegraphics[width=0.85\linewidth]{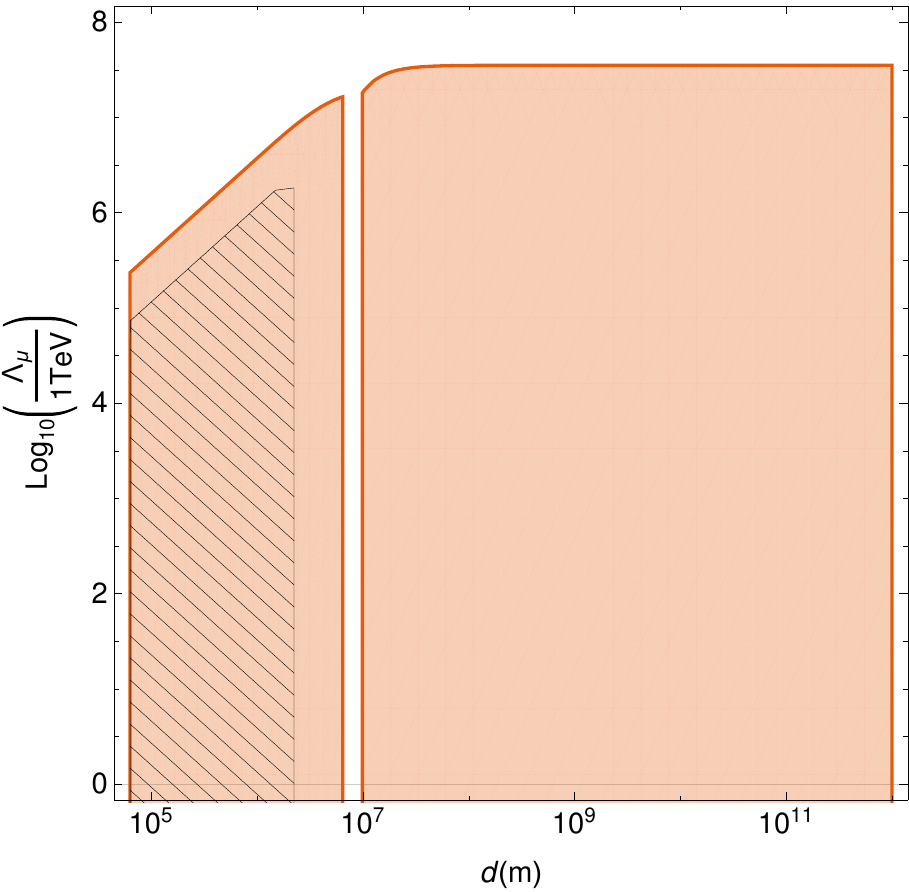}
\caption{\small{Constraints on the parameter space with the energy scale $\Lambda_\mu$ versus the size of the transient $d$, for $\mathcal{T}_\textrm{max}$ = 1 month, and assuming $\Gamma_\mu \gg \Gamma_\alpha$ in (\ref{eq:gamma_eff}). The hatched region represents the constraints previously achieved in \cite{Roberts:2017hla}. The exclusion area in orange results from the MRA introduced in this work.}\label{fig:Max-jump_Constraints_Lambda-mu}}
\end{figure}

The experimental values in (\ref{def:freq-thres_LT}) can be included in the expressions obtained for $s_*(d)$ in (\ref{eq:tilde_sab}), valid for the two ranges of transient size $d$. As expected, this leaves 3 undetermined parameters inherent to the transient model: $\Lambda_\alpha$ (or $\Lambda_\mu$), $\mathcal{T}$ and $d$. 
As a reminder, $\Lambda_\alpha$ represents the energy scale in the electromagnetic sector (\ref{def:Gamma}), while $\Lambda_\mu$ represents  the energy scale in the fermionic sector, through the proton-to-electron mass ratio as seen from equations (\ref{def:mu_ep}), (\ref{def:Gamma}) and (\ref{def:Gamma_mu}).
Fig.\ref{fig:Max-jump_Constraints} presents the new exclusion area for the parameter set ($\Lambda_\alpha$,$d$) obtained from the results of our maximum reach analysis (MRA) for the maximum encounter time $\mathcal{T}$ of 1 month. The way of interpreting our results is the following: Transient objects with a radial profile similar to Eq.~(\ref{def:DW_phi-profile}) (fulfilling the local DM density) and an electromagnetic energy scale $\Lambda_\alpha$ inferior to 10$^8$ TeV  do not exist experimentally if their size $d$ ranges between 10$^8$ and 10$^{12}$ m and if their mean period of encounter $\mathcal{T}$ is of the order of one month. The limit on the associated variation of the fine structure constant $\frac{\Delta \alpha}{\alpha}$ is of the order of 10$^{-13}$. A similar plot presented in Fig. \ref{fig:Max-jump_Constraints_Lambda-mu} has been obtained for the fermion interaction with the associated energy scale $\Lambda_\mu$. 

We have made a sharp cut-off in Fig. \ref{fig:Max-jump_Constraints} and Fig. \ref{fig:Max-jump_Constraints_Lambda-mu} for $d = 70$ km, which corresponds to the expected limit of the servo loop time. The servo loop time is the fastest disturbance, due to atomic frequency changes, that can be recorded by the Galileo clocks. Indeed, over sufficiently short intervals, all atomic clocks derive their stability from quartz oscillators. It turns out that the value of the Galileo clocks' servo loop time is not public. However, the value of GPS clocks is known to be within 0.1 s, as explained in \cite{Roberts:2017hla}. Making the realistic assumption that the value for Galileo satellites is of the same order, the servo loop problem only concerns the case of very thin domain walls. If we consider the maximum value for GPS clocks, a domain wall crossing time through satellite clock of 0.1 s represents a size $d$ of 70 km, given the galactic escape velocity. Note that \cite{Roberts:2017hla} considered sensitivity to DM events below the servo loop time, a region of parameter space where sensitivity is determined by the response of the quartz oscillator to variation in fundamental constants. However, extending such a study to our MRA with the Galileo satellite is beyond the scope of this work, which concentrates on the unconstrained region of parameter space covering large transients.

The results obtained with the MRA, i.e. the orange area in Fig.\ref{fig:Max-jump_Constraints} and Fig.\ref{fig:Max-jump_Constraints_Lambda-mu}, provide two improvements with respect to \cite{Roberts:2017hla}: (i) we extend dramatically the constraints by dealing with large transients with $d$~>~2250~km and (ii) we improve the constraints by a factor 5 for thin DW of range size [70, 2250] km. The analysis for large values of $d$ applies to various types of transients (not only DW) with a quadratic scalar coupling to matter fields. The improvement for thin DW is entirely due to the better stability of the on-board passive H-Masers.
 
 However, the general character of the hyperbolic profile as a reasonable approximation needs to be verified. This exclusion area was obtained using the clock noise but did not consider possible systematic effects that might occur during an event (although in an improbable and unfortunate manner). Indeed, all our analysis was based on a 95\% level of occurrence, but associating this percentage with a reliable confidence level is statistically uncertain due to the absence of treatment of systematics. However this exclusion area provides an important and realistic information about the types of couplings, sizes and encounter times that are unlikely to happen.

\section{Conclusions}\label{sec:Conclusion}

We have extended the toy case of thin DW introduced in \cite{Derevianko:2013oaa,Roberts:2017hla} to transient variations of fundamental constants, with event durations from a few minutes to a few months. Our simulations have shown that large transients would induce observable periodic signatures on GNSS clock phase data, with a period equal to the satellite orbital period, and a modulated amplitude.  

In this paper, we have developed a straightforward method to provide first constraints on the coupling parameters. This method is based on a maximum reach analysis (MRA). This method enables us to screen a very large region of the parameter space (especially considering the new window of large transients),  to derive preliminary constraints on the explored parameters, and furthermore to rule out event configurations for which attempting to detect transients would not be productive, as we know they are highly unlikely to occur. Our results, presented in Fig.~\ref{fig:Max-jump_Constraints}, provide a huge extension of the constraints compared to the results available in the literature. In particular, they provide the first   direct constraints ever obtained for large transients, besides those obtained indirectly through the model-dependent non-transient signatures obtained for one sign of the coupling constant and for a double-well potential as in  \cite{Stadnik:2020bfk}).

Finally, models of DM objects are one way among many others for obtaining observable variations of the fundamental constants through our experimental setup. Hence, the constraints on the parameter space for DM models could be extended to other compatible models of transient variations of fundamental constants. 

While the MRA presented here already provided significant constraints in the space parameters, a full statistical analysis based on correlations between clocks, considering the modeling as shown e.g. in Fig.~\ref{fig:sim-sig_thin-DW} and Fig.~\ref{fig:sim-sig_thick-DW},  will be presented in a future paper. This will firstly allow  putting further constraints on the coupling parameters and secondly show evidence of possible events if some correlations are statistically significant.  

\section*{Acknowledgements}
The work reported in this paper has been performed and fully funded under a contract of the European Space Agency in the frame of the EU Horizon 2020 Framework Program for Research and Innovation in Satellite Navigation. The views presented in the paper represent solely the opinion of the authors and should be considered as R\&D results not necessarily impacting the present and future EGNOS and Galileo system designs.



\bibliography{GASTON.bib}









\end{document}